# Crystal electric field parameters for $Yb^{3+}$ ion in $YbRh_2Si_2$

A S Kutuzov and A M Skvortsova

Kazan Federal University, Kremlevskaya, 18, Kazan 420008, Russian Federation

E-mail: Alexander.Kutuzov@gmail.com

**Abstract.** The tetragonal crystal electric field parameters for $Yb^{3+}$ ion in $YbRh_2Si_2$ are determined from the analysis of the literature data on angle-resolved photoemission, inelastic neutron scattering and electron paramagnetic resonance.

## 1. Introduction

$YbRh_2Si_2$ and $YbIr_2Si_2$ belong to the class of heavy-fermions Kondo-lattice compounds which attracted the interest of scientists due to the very peculiar magnetic, thermal and transport properties. The discovery of electron paramagnetic resonance (EPR) in the Kondo lattices $YbRh_2Si_2$ and $YbIr_2Si_2$ below the thermodynamically measured Kondo temperatures was very unexpected [1-3] and stimulated different approaches for its explanation [4-7]. In particular, in [6, 7] is shown that the idea of a collective spin motion of quasi-localized Yb $f$-electrons and wide-band electrons successfully explains the EPR signal existence. In [8] the static magnetic susceptibility of $YbRh_2Si_2$ and $YbIr_2Si_2$ were also successfully studied, based on entirely local properties of $Yb^{3+}$ ions in a tetragonal crystal electric field (CEF). Until recently the available experimental data were not sufficient to unambiguously determine the CEF parameters and therefore eigenfunctions of $Yb^{3+}$ ion [9] (but attempts have been made [10, 11]). Though the energy splittings between CEF states of $Yb^{3+}$ ion in $YbRh_2Si_2$ are known from inelastic neutron scattering (INS) experiments [12], but the symmetry identification of CEF levels was only recently carried out by performing angle-resolved photoemission spectroscopy (ARPES) experiments [13]. This new information together with INS and EPR data give us an opportunity to find one set of tetragonal CEF parameters for $Yb^{3+}$ ion in $YbRh_2Si_2$, as reported in details in this paper.

Besides, $Yb^{3+}$ ion was used as a probe in EPR experiments in high-temperature superconductors, in particular $YBa_2Cu_3O_x$ ($6 < x < 7$) compounds, in which yttrium was partially substituted by ytterbium [14-18]. The samples with oxygen content $x < 6.4$ have a tetragonal crystal symmetry. So, the problem of $Yb^{3+}$ ion in tetragonal CEF seems to be actual and important.

## 2. $Yb^{3+}$ ion in tetragonal CEF

A free $Yb^{3+}$ ion has a $4f^{13}$ configuration with one term $^2F$. The spin-orbit interaction splits the $^2F$ term into two multiplets: $^2F_{7/2}$ with $J = 7/2$ and $^2F_{5/2}$ with $J = 5/2$, where $J$ is the value of the total momentum $\mathbf{J} = (J_x, J_y, J_z)$. Multiplets are separated by about 1 eV [19]. As the spin-orbit coupling is much stronger than the CEF in the case of rare earth, we will consider only the ground multiplet $^2F_{7/2}$ with states $|J = 7/2, M_J\rangle \equiv |M_J\rangle$, where $M_J$ is the eigenvalue of $J_z$. If $z$ axis coincides with tetragonal axis of CEF and $x$ and $y$ axes are directed along the twofold axes then the Hamiltonian of the $Yb^{3+}$ ion interaction with the tetragonal CEF could be written via equivalent operators $O_k^q(\mathbf{J})$ [19] as follow:

$$H = \alpha B_2^0 O_2^0 + \beta(B_4^0 O_4^0 + B_4^4 O_4^4) + \gamma(B_6^0 O_6^0 + B_6^4 O_6^4), \tag{1}$$

where $B_k^q$ are the CEF parameters, $\alpha = 2/63$, $\beta = -2/1155$, $\gamma = 4/27027$ [19].

From the group theory it is known that the two-valued irreducible representation $D^{7/2}$ of rotation group contains two two-dimensional irreducible representations $\Gamma_7^t$ and $\Gamma_6^t$ of the double tetragonal group: $D^{7/2} = 2\Gamma_7^t + 2\Gamma_6^t$ [19]. Therefore the states of Yb$^{3+}$ in the tetragonal CEF are four Kramers doublets. As the decomposition of $D^{7/2}$ includes twice each of representations $\Gamma_7^t$ and $\Gamma_6^t$, the matrix of operator (1) could be expressed via two two-dimensional matrices

$$\begin{pmatrix} 2C_1 & C_3 \\ C_3 & 2C_2 \end{pmatrix} \text{ and } \begin{pmatrix} 2A_1 & A_3 \\ A_3 & 2A_2 \end{pmatrix}, \tag{2}$$

the former corresponding to bases $|5/2\rangle, |-3/2\rangle$ and $|-5/2\rangle, |3/2\rangle$, the latter corresponding to bases $|7/2\rangle, |-1/2\rangle$ and $|-7/2\rangle, |1/2\rangle$. It is convenient to introduce parameters $C$, $A$ and $D$:

$$C = C_1 - C_2 = 4B_2^0/21 + 40B_4^0/77 - 560B_6^0/429, \quad A = A_1 - A_2 = 4B_2^0/7 + 8B_4^0/77 + 80B_6^0/143, \tag{3}$$

$$D = -C_1 - C_2 = A_1 + A_2 = 2B_2^0/21 - 64B_4^0/77 - 160B_6^0/429,$$

where $C_1 + C_2 + A_1 + A_2 = 0$ as traces of $O_k^q$ are equal to zero. $C_3$ and $A_3$ are

$$C_3 = -8\sqrt{3}B_4^4/77 - 80\sqrt{3}B_6^4/1287, \qquad A_3 = -8\sqrt{35}B_4^4/385 + 80\sqrt{35}B_6^4/3003. \tag{4}$$

Let us define eigenvectors of matrices (2) $(c_{1,2}, \pm c_{2,1})$ and $(a_{1,2}, \pm a_{2,1})$ via angular parameters $\varphi_7$ and $\varphi_6$ which correspond to $\Gamma_7^t$ and $\Gamma_6^t$ symmetries:

$$c_1 = \cos(\varphi_7/2), \quad c_2 = \sin(\varphi_7/2) \quad \text{and} \quad a_1 = \cos(\varphi_6/2), \quad a_2 = \sin(\varphi_6/2). \tag{5}$$

Since matrices (2) are diagonal in the bases of their eigenvectors we can find the relations between our angular parameters and CEF parameters:

$$\tan\varphi_7 = \frac{C_3}{C} = -\frac{2}{\sqrt{3}}\left(\frac{117B_4^4 + 70B_6^4}{143B_2^0 + 390B_4^0 - 980B_6^0}\right), \quad \tan\varphi_6 = \frac{A_3}{A} = -\frac{2}{3}\sqrt{\frac{7}{5}}\left(\frac{39B_4^4 - 50B_6^4}{143B_2^0 + 26B_4^0 + 140B_6^0}\right), \tag{6}$$

it is enough to take $-\pi/2 \leq \varphi_{6,7} \leq \pi/2$. The eigenenergies $E_k$, wave functions and $g$-factors of Kramers doublets are given in table 1. In this table $^k\Gamma_7^t$ and $^k\Gamma_6^t$ are symmetry symbols, where $k = 1..4$ is the number of Kramers doublet. The arrow $\uparrow$ or $\downarrow$ in wave functions corresponds to the upper or lower sign and denotes up and down effective spin projection. They have been chosen such that $\langle\uparrow|J_+|\downarrow\rangle \neq 0$, where $J_+ = J_x + iJ_y$. Moreover, the phases of the wave function have been chosen as $\theta|\uparrow\rangle = |\downarrow\rangle$, where $\theta$ is a time reversing operator [19]. In $g$-factors left and right indexes correspond to the upper and lower signs; $g_J = 8/7$ is the Lande $g$-factor.

**Table 1.** Energies, wave functions and $g$-factors of Yb$^{3+}$ ion in tetragonal crystal electric field.

| $E_{1,2} = -D \pm C/\cos\varphi_7$ | $E_{3,4} = D \pm A/\cos\varphi_6$ |
|---|---|
| $|^1\Gamma_7^t \uparrow,\downarrow\rangle = \pm c_1|\pm 5/2\rangle \pm c_2|\mp 3/2\rangle$ | $|^3\Gamma_6^t \uparrow,\downarrow\rangle = \pm a_1|\mp 7/2\rangle \pm a_2|\pm 1/2\rangle$ |
| $|^2\Gamma_7^t \uparrow,\downarrow\rangle = \mp c_2|\pm 5/2\rangle \pm c_1|\mp 3/2\rangle$ | $|^4\Gamma_6^t \uparrow,\downarrow\rangle = \mp a_2|\mp 7/2\rangle \pm a_1|\pm 1/2\rangle$ |
| $g_\parallel(^{1,2}\Gamma_7^t) = g_J(5c_{1,2}^2 - 3c_{2,1}^2) = g_J(1 \pm 4\cos\varphi_7)$ | $g_\parallel(^{3,4}\Gamma_6^t) = g_J(a_{2,1}^2 - 7a_{1,2}^2) = -g_J(3 \pm 4\cos\varphi_6)$ |
| $g_\perp(^{1,2}\Gamma_7^t) = \mp 4\sqrt{3}g_J c_1 c_2 = \mp 2\sqrt{3}g_J \sin\varphi_7$ | $g_\perp(^{3,4}\Gamma_6^t) = -4g_J a_{2,1}^2 = -2g_J(1 \mp \cos\varphi_6)$ |

The Zeeman energy $g_J\mu_B\mathbf{HJ}$ in the basis $|\uparrow\rangle, |\downarrow\rangle$ of each doublet could be represented by matrix

$$H_{Zeeman} = g_\parallel \mu_B H_z S_z + g_\perp \mu_B (H_x S_x + H_y S_y), \tag{7}$$

where

$$g_\| = 2g_J \langle \uparrow | J_z | \uparrow \rangle, \qquad g_\perp = g_J \langle \uparrow | J_+ | \downarrow \rangle \qquad (8)$$

and **H** is the magnetic field, **S** is the effective spin operator with $S=1/2$, $\mu_B$ is the Bohr magneton, $g_\|$ and $g_\perp$ are $g$-factors when the field is applied parallel and perpendicular to the tetragonal $z$-axis, respectively (table 1).

In the case of cubic symmetry $B_2^0 = 0$, $B_4^4 = 5B_4^0$ and $B_6^4 = -21B_6^0$, so that $\tan\varphi_7 = -\sqrt{3}$, $\tan\varphi_6 = -\sqrt{35}$, $c_1 = \sqrt{3}/2$, $c_2 = -1/2$, $a_1 = \sqrt{7/12}$, $a_2 = -\sqrt{5/12}$. In accordance with expansion $\Gamma_8 = \Gamma_7^t + \Gamma_6^t$ [19] the doublets $^2\Gamma_7^t$ and $^3\Gamma_6^t$ merge into a cubic quartet $\Gamma_8$ with energy $E(\Gamma_8) = -16B_4^0/77 + 1280B_6^0/429$. The doublets $^1\Gamma_7^t$ and $^4\Gamma_6^t$ turn into cubic doublets $\Gamma_7$ and $\Gamma_6$ with energies $E(\Gamma_7) = 144B_4^0/77 - 320B_6^0/143$ and $E(\Gamma_6) = -16B_4^0/11 - 1600B_6^0/429$ and with isotropic $g$-factors $g(\Gamma_7) = 3g_J = 3.429$ and $g(\Gamma_6) = -7/3g_J = -2.667$, respectively (see figure 1). Here $\Gamma_{6,7,8}$ are irreducible representations of double cubic group [19].

As $g$-factors of each doublet depend only on one parameter $\varphi_6$ or $\varphi_7$ (table 1) we can find the equation relating $g_\|$ and $g_\perp$. Figure 1 represents the diagram of $g$-factors. The solid and dashed parts of the line $g_\| + 2g_\perp + 7g_J = 0$ correspond to the doublets $^4\Gamma_6^t$ and $^3\Gamma_6^t$, the solid and dashed parts of the ellipse $(g_\| - g_J)^2/4 + g_\perp^2/3 = 4g_J^2$ correspond to the doublets $^2\Gamma_7^t$ and $^1\Gamma_7^t$. The line and the ellipse meet in the point $(-g_J, -3g_J)$ marked by a star.

On the $g$-diagram (figure 1) we marked the experimental $g$-points which correspond to the absolute values of YbRh$_2$Si$_2$ $g$-factors measured in the EPR experiments at 5 K [1, 2] (four points with different signs of $g_\|$ and $g_\perp$). A slight difference between experimental and theoretical $g$-factors can be explained mainly by taking into account the Kondo interaction, i.e. an exchange coupling between the 4$f$-electrons of the Yb$^{3+}$ ion and conduction electrons [8]. This $g$-diagram do not allow us to choose between ground doublets $^2\Gamma_7^t$ and $^4\Gamma_6^t$ as the value of $g$-shift is unknown. But in recent ARPES experiments on YbRh$_2$Si$_2$ it was determined that the ground doublet is $^2\Gamma_7^t$ [13]. Points on the ellipse (figure 1) nearest to the experimental $g$-points ($g_\| = -0.17$, $g_\perp = \pm 3.561$) correspond to $\varphi_7 = \pm 1.2660$ ($g_\| = -0.229$, $g_\perp = \pm 3.777$). These values of $\varphi_7$ we will use in our further calculations.

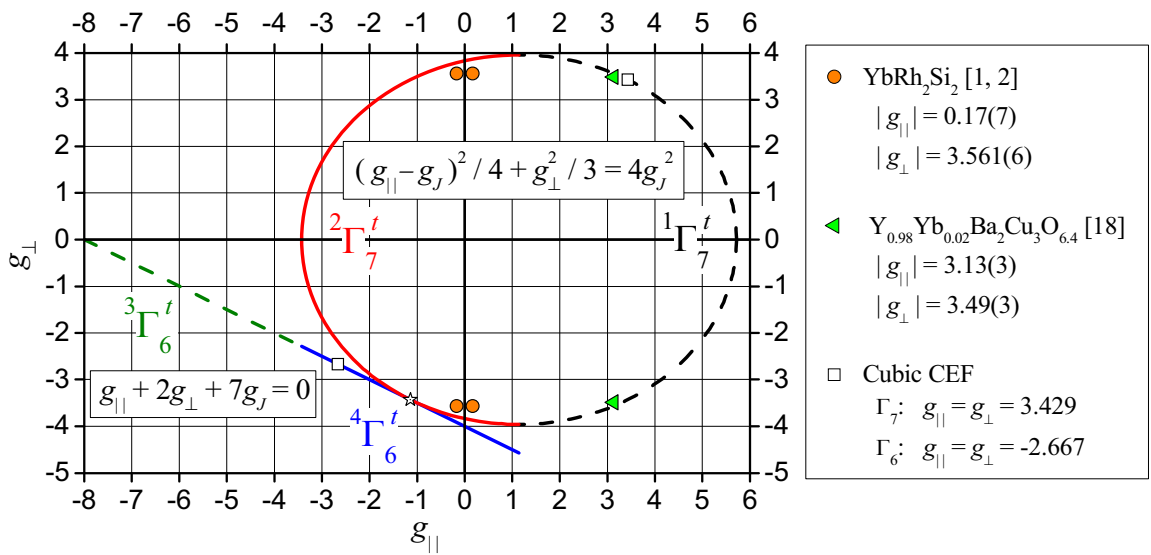

**Figure 1.** Diagram of $g$-factors of Yb$^{3+}$ ion in tetragonal crystal electric field.

### 3. Determination of CEF parameters

Let us find all possible sets of tetragonal CEF parameters for the given exited doublets energies $\Delta_1 < \Delta_2 < \Delta_3$. It follows from (3) that

$$B_2^0 = 3A/2 + C/2 + D/2, \quad B_4^0 = A/16 + 5C/16 - D, \quad B_6^0 = 39A/160 - 91C/160 - 13D/40 \quad (9)$$

and from (4) that

$$B_4^4 = -7\sqrt{35}A_3/16 - 35\sqrt{3}C_3/16, \qquad B_6^4 = 117\sqrt{35}A_3/160 - 273\sqrt{3}C_3/160. \quad (10)$$

Taking one of the doublets with energy $E_k$ (table 1) as the ground, defining the differences of doublets energies as $E_{mk} = E_m - E_k$ and solving this system of linear equations we can express $C$, $A$ and $D$ through $E_{mk}$. Substituting relations $A_3 = A \tan \varphi_6$ and $C_3 = C \tan \varphi_7$ into (10) and then $C$, $A$ and $D$ into (9) and (10) we find:

$$B_2^0 = \frac{1}{8}b + \frac{3}{4}b_6 \cos\varphi_6 + \frac{1}{4}b_7 \cos\varphi_7,$$

$$B_4^0 = -\frac{1}{4}b + \frac{1}{32}b_6 \cos\varphi_6 + \frac{5}{32}b_7 \cos\varphi_7, \qquad B_4^4 = -\frac{7\sqrt{35}}{32}b_6 \sin\varphi_6 - \frac{35\sqrt{3}}{32}b_7 \sin\varphi_7, \quad (11)$$

$$B_6^0 = -\frac{13}{160}b + \frac{39}{320}b_6 \cos\varphi_6 - \frac{91}{320}b_7 \cos\varphi_7, \qquad B_6^4 = \frac{117\sqrt{35}}{320}b_6 \sin\varphi_6 - \frac{273\sqrt{3}}{320}b_7 \sin\varphi_7,$$

where $b$, $b_6$ and $b_7$ are determined in table 2. To use (11) we have to choose the ground doublet and the exited doublets sequence to express energy differences $E_{mk}$ in table 2 through experimental values $\Delta_1 < \Delta_2 < \Delta_3$. Angular parameters $\varphi_6$ and $\varphi_7$ can take the values $-\pi/2 \leq \varphi_{6,7} \leq \pi/2$ independently, the energy scheme does not depend on them. Formulas (11) are suitable for $Yb^{3+}$ ion in any tetragonal CEF. Now we apply (11) for $Yb^{3+}$ ion in $YbRh_2Si_2$.

**Table 2.** $b$, $b_6$ and $b_7$ in (11).

| Ground doublet | $b$ | $b_6$ | $b_7$ |
|---|---|---|---|
| $^1\Gamma_7^t$ | $E_{31} - E_{21} + E_{41}$ | $E_{31} - E_{41}$ | $-E_{21}$ |
| $^2\Gamma_7^t$ | $E_{32} - E_{12} + E_{42}$ | $E_{32} - E_{42}$ | $E_{12}$ |
| $^3\Gamma_6^t$ | $E_{43} - E_{13} - E_{23}$ | $-E_{43}$ | $E_{13} - E_{23}$ |
| $^4\Gamma_6^t$ | $E_{34} - E_{14} - E_{24}$ | $E_{34}$ | $E_{14} - E_{24}$ |

According to INS experiments on $YbRh_2Si_2$ [12] the excited doublets energies approximately equal to $\Delta_1 = 17$, $\Delta_2 = 25$, $\Delta_3 = 43$ meV relative to the ground doublet. The Kramers doublets sequence was recently determined from ARPES experiments [13]: $^2\Gamma_7^t$, $^4\Gamma_6^t$, $^1\Gamma_7^t$, $^3\Gamma_6^t$. Above we found $\varphi_7$ from experimental $g$-factors of $^2\Gamma_7^t$ ground doublet. So, we need only the value of $\varphi_6$ to find CEF parameters using formulas (11).

Let us consider INS experiments [12] in more detail. Magnetic part of INS spectrum of $YbRh_2Si_2$ at $T = 1.5$ K [12] is presented on figure 2 (we have digitized data from Fig. 2 of [12]). Three peaks correspond to the transitions between ground doublet and excited doublets. In [12] magnetic response has been fitted quite well by a sum of three Lorentzians. We have also fitted the data from [12] by formula used in [20-22]:

$$S_{mag}(\varepsilon) \propto \frac{\varepsilon}{1 - \exp(-\varepsilon/kT)} \sum_{i=1,2,3} I_i \frac{1 - \exp(-\Delta_i/kT)}{\Delta_i} F(\varepsilon - \Delta_i, L_i), \quad (12)$$

where $\varepsilon$ is the energy transferred from the neutron to the sample, $I_i$ are the intensities of the CEF transitions, $L_i$ are the half-width of Lorentzians

$$F(\varepsilon, L) = \frac{1}{\pi} \frac{L}{L^2 + \varepsilon^2} . \qquad (13)$$

The solid line in figure 2 indicates the fit, the fitted values of $\Delta_i$, $L_i$ and $I_i$ are given in the second row of table 3.

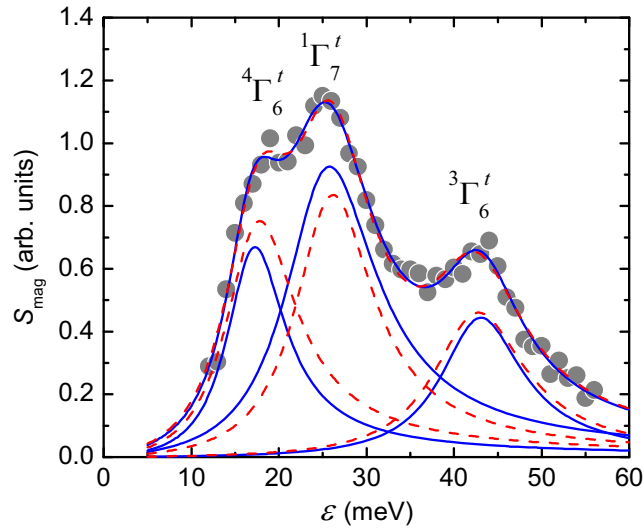

**Figure 2.** Magnetic part of the inelastic neutron scattering spectrum of YbRh$_2$Si$_2$ at $T = 1.5$ K (digitized data from Fig. 2 of [12]). The solid line indicates a fit by expression (12) with $\Delta_i$, $L_i$ and $I_i$ as independent parameters (see the second row in table 3). The dashed line indicates a fit by (12) with $\Delta_i$, $L_i$ and $\varphi_6$ as independent parameters and with $I_i$ calculated by using (15) (see the third row in table 3).

**Table 3.** Kramers doublets energies $\Delta_i$ (meV), half-width of Lorentzians $L_i$ (meV), CEF transition intensities $I_i$, angular parameter $\varphi_6$ for YbRh$_2$Si$_2$ from the data fits according to expression (12) as shown in figure 2; angular parameter $\varphi_7$ from the $g$-factors. Second and third rows correspond to solid and dashed lines in figure 2.

| $\Delta_1$ | $\Delta_2$ | $\Delta_3$ | $L_1$ | $L_2$ | $L_3$ | $I_1$ | $I_2$ | $I_3$ | $\varphi_6$ | $\varphi_7$ |
|---|---|---|---|---|---|---|---|---|---|---|
| 16.8 | 25 | 42.7 | 4.1 | 6.3 | 5.7 | 0.47 | 1 | 0.44 | | |
| 17.1 | 25.6 | 42.3 | 4.9 | 5.4 | 6.4 | 0.63 | 0.78 | 0.51 | 0.5361 | −1.2660 |

In accordance with [20-22] for polycrystalline sample or powder the intensity of transition from ground doublet $\Gamma$ to excited doublet $\Gamma'$ is given by

$$I_{\Gamma'} \propto \sum_{\alpha\sigma\sigma'} |\langle \Gamma\sigma | J_\alpha | \Gamma'\sigma' \rangle|^2 , \qquad (14)$$

where $\alpha = x, y, z$ and $\sigma, \sigma' = \uparrow, \downarrow$. For transitions from $^2\Gamma_7^t$ ground doublet to excited doublets $^4\Gamma_6^t$, $^1\Gamma_7^t$ and $^3\Gamma_6^t$ we obtain

$$I_{{}^4\Gamma_6^t} = (\sqrt{7}c_2 a_2 + \sqrt{15}c_1 a_1)^2 = I(\varphi_7)\cos^2\left(\frac{\varphi_6 - \gamma(\varphi_7)}{2}\right) \equiv I_1,$$

$$I_{{}^1\Gamma_7^t} = 32 c_1^2 c_2^2 + 12(c_2^2 - c_1^2)^2 = 4(3 - \sin^2\varphi_7) \equiv I_2, \qquad (15)$$

$$I_{{}^3\Gamma_6^t} = (\sqrt{7}c_2 a_1 - \sqrt{15}c_1 a_2)^2 = I(\varphi_7)\sin^2\left(\frac{\varphi_6 - \gamma(\varphi_7)}{2}\right) \equiv I_3,$$

where

$$I(\varphi_7) = 11 + 4\cos\varphi_7, \qquad \gamma(\varphi_7) = 2\arctan\left[\sqrt{7/15}\tan(\varphi_7/2)\right]. \qquad (16)$$

These intensities as the functions of still undefined parameter $\varphi_6$ are plotted in figure 3. Experimentally observed relation $I_3 < I_1 < I_2$ takes place for $\varphi_6$ approximately lying within the interval $0.25 < \varphi_6 < 0.65$. The fitted values of intensities are in the ratio $I_1 : I_2 : I_3 = 0.47 : 1 : 0.44$ (see the second row in table 3). The calculation of intensities using (15) gives the nearest (but not near enough) result with $\varphi_6 = 0.6155$: $I_1 : I_2 : I_3 = 0.75 : 1 : 0.71$.

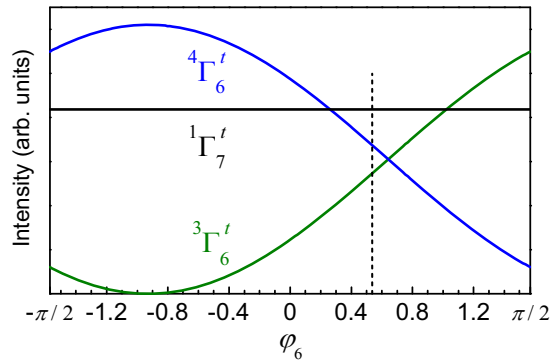

**Figure 3.** Intensities of the CEF transitions in the INS spectrum calculated by using (15) with $\varphi_7 = -1.2660$. The vertical line correspond to $\varphi_6 = 0.5361$.

We made also another fit using formula (12) with fitting parameters $\Delta_i$, $L_i$ and $\varphi_6$ (see the third row in table 3). In (12) the intensities $I_i(\varphi_6)$ were calculated by (15) assuming $\varphi_7 = -1.2660$. The experimental data was also very well fitted (see dashed line in the figure 2). The fitted value of $\varphi_6 = 0.5361$ is marked on figure 3 by vertical line.

We took obtained in this fitting procedure values of $\Delta_i$ and $\varphi_6$, determined earlier value of $\varphi_7$ and then we calculated CEF parameters with the help of formulas (11). Corresponding CEF parameters are presented in table 4.

**Table 4.** Crystal electric field parameters $B_k^q$ (meV) for $Yb^{3+}$ ion in $YbRh_2Si_2$.

| $B_2^0$ | $B_4^0$ | $B_4^4$ | $B_6^0$ | $B_6^4$ |
|---|---|---|---|---|
| 22.4 | −6.6 | ±29.6 | −2.3 | ±63.9 |

If we simultaneously change the signs of $\varphi_6$ and $\varphi_7$ the intensities (15) stay the same but $B_4^4$ and $B_6^4$ (11) change the signs. In table 4 positive signs of $B_4^4$ and $B_6^4$ correspond to the $\varphi_6$ and $\varphi_7$ given in table 3 whereas negative signs of $B_4^4$ and $B_6^4$ correspond to $\varphi_6$ and $\varphi_7$ with the opposite signs. So, used INS experimental data do not allow to determine the signs of $B_4^4$ and $B_6^4$. Unfortunately, the relation $g_\perp \sim \sin \varphi_7$ (see table 1) also do not help us to choose the signs of $B_4^4$ and $B_6^4$. The reason is that in usual EPR experiments only the absolute values of *g*-factors are defined, therefore we have to consider two points on *g*-diagram (figure 1) with opposite sings of $g_\perp$ and $\varphi_7$.

In the papers [10, 11] the sets of CEF parameters for $Yb^{3+}$ ion in $YbRh_2Si_2$ were found by the least squares fit to the experimental energy levels and *g*-factors (these sets of CEF parameter could be obtained from (11), see details in [9]). But these sets of CEF parameters do not agree with other experimental data. Though one set of CEF parameters from [10] agrees with doublets sequence established in ARPES experiments but gives almost zero intensity of INS transitions between ground doublet $^2\Gamma_7^t$ and excited doublet $^3\Gamma_6^t$, i.e. this set reproduces only two peaks instead of three ones experimentally observed in INS experiment (figure 2). Another set of CEF parameters from [10] and CEF parameters from [11] correspond to the ground doublet $^4\Gamma_6^t$ instead of $^2\Gamma_7^t$ in a contradiction with ARPES findings.

**4. Summary**
We analytically calculated the tetragonal CEF splitting of the ground $^2F_{7/2}$ multiplet of $Yb^{3+}$ ion. Four Kramers doublets energies and doublets wave functions are expressed via five CEF parameters. We also analytically solved the inverse problem, i.e. all possible sets of tetragonal CEF parameters that satisfy the given experimental energy scheme of $Yb^{3+}$ ground multiplet $^2F_{7/2}$ are found (see (11)). The diagram of possible *g*-factors of $Yb^{3+}$ ion in tetragonal CEF is plotted (figure 1).

We determined the values of CEF parameters for $Yb^{3+}$ ion in $YbRh_2Si_2$ (table 4) using energy intervals and intensities of CEF transitions between Kramers doublets taking from fitting of INS data [12], the Kramers doublets sequence obtained from ARPES experiments [13] and the *g*-factors found from EPR experiments [1, 2].

**Acknowledgments**
This work was supported by the Volkswagen Foundation (I/84689), and partially by the Swiss National Science Foundation IZ73Z0_128242/1 and by the Ministry of Education and Science of the Russian Federation. We wish to acknowledge professor B.I. Kochelaev for bringing our interest to this problem and continuous interest to the results.